   \font\blank=cmss10   \def\x{{\blank x}}
\def\0{\over}           \def\6{\partial}
\def\ln{\hbox{ln}}      \def\P{ {\mit\Pi} }
\def\eq#1{(\ref{#1})}   \def\ov#1{\overline{#1}}
\def\cl#1{{\cal #1}}    \def\schl#1{\widetilde{#1}}
\def\dis{\displaystyle} \def\Q{{\dis\cl Q}}
\def\be#1{\begin{equation}\label{#1}}
\def\ee{\end{equation}}
   \let\b=\beta  \let\d=\delta  \let\l=\lambda
   \let\e=\varepsilon   \let\thq=\theequation
\def\({\left(}  \def\){\right)}  \def\rki{\,\right]}
\def\lk{\,\left[\,}  \def\rk{\,\right]\,}
\def\sb{\;\hbox{\vrule height 10pt depth 3pt width .6pt}
   \lower 5pt\hbox{$\!\>$}}
\def\self{\hspace{.2cm}\raise 2pt\hbox{$
   \bigcirc$}\hspace{-.5cm}  \hbox{\vrule depth 1pt
   height -.6pt width .58cm}\hbox{$\,$}}
\def\parag#1{ \vspace{1.5cm} \hspace{.08cm}
    \parbox{14cm}{{\bf #1} \vspace{.9cm} }
    \nopagebreak[4] \indent }
\def\under#1{$\lower 0.2pt\hbox{$\;
    \underline{\vphantom{\int}\hbox{#1}}\;$}$}

\documentstyle[12pt,twoside]{article}
 
\begin{document}


    \begin{titlepage}  \begin{flushleft}   
DESY \ 97--059 \\
ITP--UH $\; 08 \, / \, 97$ \\
hep-ph/9703414 \end{flushleft} \vspace{-2cm}
   \begin{flushright}  March 1997 \hspace{1.6cm} $ $ \\
              Phys. Lett. {\bf B 404} (1997) 291 $ $ \\
   \end{flushright} \vfill \vskip 2.5cm 
   \begin{center} {\Large \bf 
Resummation of $\phi^4$ free energy \\[10pt]  
up to an arbitrary order} \\  
\vfil  \vspace{2cm} 
\renewcommand{\thefootnote}{\fnsymbol{footnote}} 
  {\large 
Jens Reinbach \ and \ Hermann Schulz
   \footnote[1]{~Electronic addresses : \
reinbach@itp.uni-hannover.de , \ 
hschulz@itp.uni--hannover.de } }
   \renewcommand{\thefootnote}{\arabic{footnote}} 
   \\[6pt] \bigskip  {\sl    
Institut f\"ur Theoretische Physik,
Universit\"at Hannover \\
Appelstra\ss e 2, D-30167 Hannover, Germany \\  }
    \vspace{3.5cm} \vfill  {\bf    
Abstract \quad }  \end{center} \begin{quotation}  \ \  
   The consistency condition, which guarantees a well
organized small--coupling asymptotic expansion for the
thermodynamics of massless $\phi^4$--theory, is generalized
to any desired order of the perturbative treatment. Based
on a strong conjecture about forbidden two--particle
reducible diagrams, this condition is derived in terms of
functions of four--momentum in place of the common toy mass
in previous treatments. It has the form of a set of gap
equations and marks the position in the space of these 
functions at which the free energy is extremal.

    \vspace{.3cm} \noindent PACS numbers : \ 
11.10.Wx \ and \ 11.15.Bt      \end{quotation}
\end{titlepage}


\let\dq=\thq \renewcommand{\theequation}{1.\dq}    
\setcounter{equation}{0}          

\parag{1. \ Introduction}

Thermal field theory \cite{kapu,bell,LW} must take care
of infrared singularities which infect a naive diagrammatic
expansion. But the way out is well known \cite{GPY}, namely
a reorganization of the perturbation series by introducing
some toy mass term to be added and subtracted in the
Lagrangian, one term becoming part of the bare Lagrangian
and the other being treated as perturbation.

At finite temperature, the convenience of first choosing
a suitable effective free Lagrangian became apparent in
studies of spontaneously broken symmetry \cite{wein,AE} and
was then encouraged by the success of the Braaten--Pisarski
resummation \cite{BP} in understanding the quark--gluon
excitations. In particular, to treat hot $\phi^4$ theory,
the toy mass was given the value of the dynamically
generated thermal mass \cite{GPY,par}. This
choice was seen to work well up to three--loop order
\cite{fstcp,az}, while in gauge theories the static limit
(Matsubara frequency $P_0=0$) of the dynamical mass was 
found to be sufficient \cite{az,paz,zk}. Even $\phi^4$
thermodynamics can be formulated using the zero--mode 
propagator \cite{pasi}. The asymptotic expansion for the
$\phi^4$ pressure is now known up to the $g^5$ term and well 
confirmed by calculations \cite{bn} using the dimensional 
reduction method. The proposal to use a gap equation for 
''an even better choice'' of the toy mass can be found in 
the book of Le Bellac \cite{bell} (\S~4.1 there), and was 
recently discussed in relation with the large--N limit and 
numerics \cite{kpp}.

There remain questions particularly about the toy mass,
as to which choice (given an accuracy of treatment) is
required according to which principle. There might be an
underlying structure the so far used mass versions are
only special cases thereof. In this note we are in search 
of this general structure. Massless $\phi^4$ theory turns
out to be sufficiently simple for studying the free
energy up to an arbitrary high order of interest.

The idea, basic to this paper, came into mind while 
re--examining the $\phi^4$ part of the paper of Arnold
and Zhai \cite{az}, whose analysis extends to $g^4\,$.
Leaving the toy mass $m$ variable, and before evaluating
thermal sum--integrals, their result for the pressure 
$-F/V$ reads
\be{1az}
 -\, {1\0V}\, F\; = \; a + {1\02} m^2 b - {1\08} g^2 b^2 
 + {1\04} \(m^2-{1\02}g^2 b \)^2 c \; + \;
 {\rm const}_m \;\; ,
\ee 
where $b = \sum G_0$ and $c=\sum G_0^2$ are functions
of $m$, as also is $a$ with the property $\6_m a
= - m b\,$. In our notations (see \S 2) $G_0
= 1/(m^2-P^2)\,$. What one could learn from \eq{1az},
first time and at low order, is as follows. The $m$ value
at which the pressure becomes minimal ($m = g
  \sqrt{b/2\,}^{\hbox to0.2pt{\hss$
  \vrule height 2.5pt width 0.6pt depth -.5pt $}\;\!}$
from $\6_m F = 0$) precisely equals the position where
the unwanted $c$ term vanishes. The suppression of the
$c$ term is in fact ''necessary to get a well--behaved
expansion in $g\,$'' \cite{az}, because otherwise $g^3$
terms would arise from perturbative $g^4$ diagrams. Note
that $c = - {1\02m}\6_m b$, which becomes $\sim 1/g\,$
through evaluation. In short, the extremum condition for
the free energy agrees with the consistency condition
\ --- \ at least in the case \eq{1az} at hand.

The requirement for generalization of this agreement
makes the plan for the present paper. We shall search for
the extremum condition of the free energy, developed up to
a given order $g^{2\l}$ of perturbation expansion. Then,
we read this condition tentatively as the consistency
condition. The latter is then shown to remove all
unwanted diagrams, whose two or more lines at the same 
inner momentum would violate a well--organized asymptotic 
expansion (see \eq{5con} below). There might be no other 
mechanism ruining the expansion. But this, admittedly, we 
are only able to state as (strong) conjecture.

At first glance, the Lagrangian we shall work with,
\be{1la}
   \cl L \; = \cl L _0 + \cl L _{\rm int} \; = \;
   {1\02} \( \6 \phi \)^2 - {1\02} \phi \( Y \phi \)
   - {g^2\04!} \phi^4 + {1\02} \phi \( \ov Y \phi \) 
  \;\; ,
\ee 
is nothing but the usual one. Just $Y$ appears in place
of the squared toy mass $m^2$. Soon, however, we shall
generalize $Y$ to be an even function of four momentum
(see \eq{2yp} below). Let the bar over the second $Y$ in
\eq{1la} be a reminder that it is a part of the 
interaction (imagine $\ov Y \propto g^2\,$, but even higher
powers will be included in \S~2). Of course, $\ov Y $ must
be set equal to $Y$ at an appropriate stage of analysis
in order to reinstall the original theory. The original 
theory is massless. At a second glance one could miss 
counter terms in \eq{1la}. However, our analysis will 
be mainly diagrammatic, hence making renormalization
details more suitable for working in afterwards.

   The toy function $Y$, being part of $\cl L_0$ in the 
above Lagrangian, is nothing but a variable self--energy
in the bare propagator $G_0 (P) = 1/(Y(P)-P^2)$.
Note that this generalization opens the door for using
functional methods. So, let us state here also the most 
useful technical detail, which is the relation
\be{1kap}
    2 \, G_0^2(P) \;\d_{G_0 (P)} \;\ln \( Z \) 
    = 2 \b \;\d_{Y(P)}\; F \;
    {\vrule depth 7pt height 10pt width .2pt}_{\;\ov Y \;
    {\rm held\; fixed}}  \; = \; G (P) \;\; ,
\ee 
where Z is the partition function, $F=-T \ln (Z)$, $T=1/\b$
the temperature and $G$ the exact propagator to the
Lagrangian \eq{1la}. Any wisdom on the free energy
corresponds to one on the Greens function and vice versa.
We learned about this connection from the text book of
Kapusta \cite{kapu} (equation (3.24) there, or (4.2.5) in
\cite{LW}). But it goes back to the sixties \cite{bloch}, 
at least.

Section 2 collects our notations and a few diagrammatic 
details. In Section 3 the extremum of $F$ is found for
a general order $g^{2\l}$ at which the perturbation series
is truncated. Then, in Section 4, the extremum condition
is shown to remove all unwanted, i.e. $g$--order--reducing 
diagrams. Simple diagrammatic rules are given which
determine the resummed series. The way it reproduces the
known results is detailed in Section 5, followed
by conclusions in Section 6. Some hints for generating
diagrams and combinatoric factors are found in the 
Appendix. 


\let\dq=\thq \renewcommand{\theequation}{2.\dq}    
\setcounter{equation}{0}          
   
\parag{2. \ The perturbation series}

There are ''red'' $g$'s in the Lagrangian \eq{1la},
those in the interaction part (last two terms), and
''blue'' ones in the bare Lagrangian. The latter
remain subordinate parameters up to the step where
the functions $Y$ and $\ov Y$ are identified. Then
they modify the perturbation expansion through
the evaluation of sum--integrals. The order $n$ of a
diagram (if it carries the prefactor $g^{2n}$) is 
determined by red $g$'s. We work with the Matsubara 
contour and Minkowski metrics ($+\!$ $-\!$ $-\!$ $-$).
Four--momenta $P$ have the components $(\, i\,2\pi n T\, ,
\, {\bf p}\, )\,$, and $x=( -i\tau\, ,\, {\bf x}\, )\,$.

Mass terms are made momentum dependent conveniently
in Fourier space. Especially, the object $(Y\phi)$ in 
\eq{1la} reads 
\be{2yp}
   \( Y \phi \) = \int^\b_{x'} \sum_P e^{-iPx} e^{iPx'} 
   Y(P) \,\phi ( x' ) = \sum_P e^{-iPx} Y(P)\,
   \schl \phi (P) \;\; .
\ee 
Variable ''masses'' of the above type are familiar
(in the more recent thermal field theory) from the
formulation of effective actions \cite{effa,nbag,fr}
and have been studied recently in another context 
\cite{vari}. Our notations are in essence those of 
\cite{vari}. The following four symbols, the first two
being used in \eq{2yp} above, may need explanation$\,$,
\be{2sym}
  \int_x^\b \equiv \int^\b_0 \! d\tau \int\! d^3r \;\; ,
  \;\; \sum_P \equiv {1\0\b V} \sum_n \sum_{{\bf p}} \;\;
  , \;\; \sb_P \equiv \b V \d_{\schl \jmath (P)} \;\; ,\;\;
  \lk P \rk \equiv \b V \d_{n,0} \d_{{\bf p} , 0} \;\; ,
\ee 
while our Fourier convention might be obvious from
\eq{2yp}. The harmony between the above four definitions 
becomes apparent in the following relations :
\be{2rel}
   \int_x^\b e^{iPx} = \lk P \rk \quad , \quad
   \sum_K \lk K-P \rk = 1  \quad , \quad
   \sb_P \,\sum_K \schl \phi (K) \,\schl\jmath (K)
   = \schl \phi (P)  \;\; .
\ee 
Analyzing a diagram, there is always one thermal Kronecker
of zero argument left, giving $\lk 0 \rk = \b V$. The
rules for going to continuous three--momenta are
$ \sum_{{\bf p}} \to V \int\! d^3p\, /(2\pi)^3$ and
$ V \d_{{\bf p} , 0} \to (2\pi)^3\d({\bf p})\,$.

Rewriting the partition function $Z=\hbox{Tr}(e^{-\b H})$
into the functional integral language (and removing 
multiple vacuum--to--vacuum amplitudes) is a standard 
procedure \cite{kapu}, resulting in
\be{2z}
   Z = Z_0 \cdot Z_{\rm int} \quad , \quad
   \ln \( Z_{\rm int} \) = \lk \; \( e^\Q - 1 \) \; W_0 \; 
   \rki_{j=0 \; ,\;\hbox{\scriptsize connected} } \;\; .
\ee 
We have reasons to supply \eq{2z} with details. The 
operator $\Q $ in \eq{2z} is a sum of two$\,$:
\be{2qy}
  \Q \;\; = \;\;\Q_Y \; +\; \Q_g \qquad\hbox{with}\qquad
  \Q_Y = {1\02} \,\sum_P \ov Y (P) \;\sb_P \,\sb_{-P}
\ee 
\be{2qg}
   \hbox{and} \qquad
   \Q_g = -{g^2\0 4!} \,\sum_{P_1 P_2 P_3 P_4}
   \lk P_1 + P_2 + P_3 + P_4 \rk \,
   \sb_{P_1} \,\sb_{P_2} \,\sb_{P_3} \,\sb_{P_4} \;\;\; .
\ee 
They act on the functional
\be{2w}
  W_0 = e^{ {1\02} \sum_P \schl \jmath (-P)\, G_0 (P)\,
        \schl \jmath (P) }  \quad , \quad 
        G_0 (P) = { 1 \;\; \0 Y(P) - P^2 } \;\; .
\ee 
Finally, the bare partition function $Z_0$ is given by
\be{2bare}
  Z_0 = \cl N \int\! \cl D \schl \phi \; e^{ {1\02} 
  \sum_P \schl \phi (-P) \; \hbox{${-1\0 G_0(P)}$}\;
  \schl \phi (P) }
\ee 
with the functional measure $\cl N$ to be determined
such that, with $Y \to 0\,$, $Z_0$ turns into the 
partition function of blackbody radiation with only 
one ''polarization'' possible \cite{vari}, see also
\eq{4lim} below. Note that, with \eq{2z}, everything
about diagrams of any order is cast into one line.
Some general properties of the perturbation series,
as will be seen, are better distilled from this line
than by studying a variety of diagrams.

As a first step into the diagrammatic analysis let 
us define the self--energy $\P$ through Dyson's equation
$G=G_0 + G_0\; [\,\ov Y - \P\,]\, G$. Thus, the
exact propagator may be written as
\be{2g}
  G(P)\; = \; { 1\0 Y(P) - P^2 - \lk \ov Y(P)
           - \P (P)\rk } \; = \; G_0 + G_0^2 \lk \; \rk 
           + G_0^3 \lk \;\, \rki^2 + \ldots
\ee 
Equating $Y=\ov Y$ shows that $\P$ is the true (exact) 
self--energy of the theory to be studied. On the other
hand, $\P - \ov Y$ could be named the ''perturbative''
self--energy.

We shall have to specify the $n$-th order term of $G\,$.
Red $g$'s do occur squared only. Let an index $n$ (as 
well as the term ''$n$-th order'') refer to the prefactor
$g^{2n}$ (red $g$'s). Then, from \eq{2g}, if $\P \sim 
\ov Y \sim g^2$, we would have $G_n = G_0^{n+1} \lk
\ov Y -\P \rki^n$. But this is not true, since
there are higher orders in $\P$. This suggests including 
higher orders in $\ov Y$ and $Y$, too$\,$:
\be{2ser}  
   \P = \P_1 + \P_2 + \P_3 +  \ldots \qquad , \qquad
   \ov Y = \ov Y_1 + \ov Y_2 + \ov Y_3 +  \ldots 
\ee 
The Lagrangian \eq{1la}, $\cl L = \cl L_0
+ \cl L_{\rm int}\,$, is now fully specified.

The $n$-th order contribution 
$f_n = - \b F_n = \ln\( Z_{\rm int} \)_n$ to the free 
energy is obtained from the following diagrammatic rules
(formulated as near to \cite{kapu} as possible)$\,$:
\be{2rules}       \hspace{.2cm}
   \parbox{14cm}{{\footnotesize
1. Draw all connected diagrams of $n$-th order. Lines may
   carry crosses of order $m\,$.   
   \hfill \\[-1pt]
2. Determine the combinatoric factor for each diagram.
   \hfill \\[-1pt]
3. Label the lines with momenta, conserved at vertices,
   and associate a $G_0$ which each.  
   \hfill \\[-1pt]
4. There is a factor $-g^2/4!$ at each four vertex, and a
   factor ${1\02}\ov Y_m$ at a cross of order $m$.
   \hfill \\[-1pt]
5. Sum over momenta with the symbol \eq{2sym},
   and put a factor $\,[\, 0\,]\, =\b V$ in front of all
   this. }}
\ee 
The weak point is in rule 2. How to determine combinatoric 
factors$\,$? We answer this question in Appendix A. Let 
figure 1 illustrate what the first two of the rules 
\eq{2rules} bring about. 

%
%
\begin{figure}[bht] \unitlength1cm 
    \begin{picture}(14.7,3.4)
\put(.2,2.4){$f_3 = -\b F_3 \;\; = \;\; 2\cdot 144$}
   \put(4.7,2.5){\circle{.4}\circle{.4}} 
   \put(4.9,2.84){\circle{.4}}
\put(5.5,2.4){$+\;\; 3\cdot 144 $} 
\put(7.6,2.56){\circle{.4}\circle{.4}\circle{.4}\circle{.4}}
\put(9.2,2.4){$+\;\; 2\cdot 144$}
  \put(11.4,2.56){\circle{.4}} \put(11.15,2.4){\circle{.2}}
  \put(11.65,2.4){\circle{.2}} \put(11.4,2.86){\circle{.2}}
\put(12,2.4){$+\;\; 4\cdot 144$}
   \put(14,2.56){\circle{.2}} \put(14.3,2.56){\circle{.4}}
   \put(14.5,2.56){\circle{.4}}
\put(3,1.3){$+\;\; 144 $} 
  \put(4.6,1.46){\circle{.4}\circle{.4}\circle{.4}}
\put(6,1.3){$+\;\; 144$}
  \put(7.6,1.46){\circle{.4}\circle{.4}\circle{.4}}
\put(9.2,1.3){$+\;\; 96$}
  \put(10.6,1.46){\circle{.4}} \put(10.8,1.46){\circle{.4}}
\put(11.6,1.3){$+\;\; 12$}
  \put(13.4,1.46){\circle{.4}\circle{.4}}
\put(4.92,1.19){\x}   \put(4.84,.98){{\tiny (1)}} 
\put(8.49,1.38){\x}   \put(8.69,1.38){{\tiny (1)}} 
\put(10.92,1.38){\x}  \put(11.12,1.38){{\tiny (1)}} 
\put(13.13,1.38){\x}  \put(12.75,1.38){{\tiny (1)}} 
\put(13.92,1.38){\x}  \put(14.12,1.38){{\tiny (1)}} 
\put(3,0){$+\;\; 24$}
  \put(4.4,.16){\circle{.4}} \put(4.88,.16){\circle{.6}}
\put(5.9,0){$+\;\;\dis {4\03}$} \put(7.3,.16){\circle{.6}} 
\put(8,0){$+\;\; 12$} \put(9.4,.16){\circle{.4}\circle{.4}}
\put(10.7,0){$+\;\; 2$}    \put(12,.16){\circle{.6}} 
\put(12.7,0){$+\;\; $}     \put(13.7,.16){\circle{.6}}
\put(5.05,.26){\x}   \put(5.26,.26){{\tiny (1)}} 
\put(5.06,-.08){\x}  \put(5.26,-.08){{\tiny (1)}} 
\put(7.2,.36){\x}    \put(7.12,.6){{\tiny (1)}} 
\put(7,-.1){\x}      \put(6.92,-.3){{\tiny (1)}} 
\put(7.45,-.1){\x}   \put(7.37,-.3){{\tiny (1)}} 
\put(9.91,.08){\x}   \put(10.13,.08){{\tiny (2)}} 
\put(11.92,.36){\x}  \put(11.84,.6){{\tiny (2)}} 
\put(11.92,-.2){\x}  \put(11.84,-.4){{\tiny (1)}} 
\put(13.62,.36){\x}  \put(13.54,.6){{\tiny (3)}} 
\end{picture} \caption[f1]{\label{f1}{\footnotesize
 All third order contributions ($g^6$) to the free energy
 including their combinatoric factors. The perturbative
 order of mass term insertions is indicated in parentheses.
 The only two--particle irreducible diagrams are the first
 and the last one.  }} \vspace{.5cm} \end{figure}
%


\let\dq=\thq \renewcommand{\theequation}{3.\dq}    
\setcounter{equation}{0}          
   
\parag{3. \ Varying $F$ with respect to $Y$}

After $\ov Y$ is identified with $Y$, this function drops 
out in the Lagrangian. Hence, it can not survive in the
exact free energy$\,$: $\d_Y F^{\rm exact}=0$. If,
however, the perturbation series is truncated at some 
order $\l$ (retaining $g^{2\l}$, neglecting $g^{2\l + 2}$,
red $g$'s), the perturbative object $F^{(\l )}\,$ 
{\sl does}$\,$ depend on $Y$, even after $Y=\ov Y$ has 
been worked in (let $F^{(\l )}\,$ be defined as the
result of this setting). Note the mystery. When varying
$F^{(\l )}$ with respect to $Y_1=\ov Y_1\, ,\;\ldots\, 
,\; Y_\l=\ov Y_\l$, we do something that has no
counterpart on the exact side. In particular, the $Y$'s
are {\sl not} parameters measuring the departure from
equilibrium. So, the free energy must not exhibit its
minimum property. In fact, varying toy parameters, it
may have a maximum \cite{lutt}, as it happens with the
low--order example \eq{1az}, indeed.

   There is a nice exercise which checks the above
statement as well as the setup of \S~2$\,$: verify the
non--$Y$--dependence of the exact $F$ by explicit
calculation. With $\cl E$ an operator that sets
$\ov Y = Y\,$ (''the equalizer'') we may do so by
\be{3ex}
  \d_{Y(P)} \;\cl E \; F  = \cl E \,\(
  \d_Y\, F + \d_{\ov Y}\, F \) \; =\; \cl E \, {T\02} G
   - T \cl E \,\lk \( \d_{\ov Y}\, \Q_Y \)
     e^\Q W_0 \,\rki_{j=0,\;{\rm connected}} \;\; .
\ee 
The first term to the right is obtained from \eq{1kap}.
For the second term, we used \eq{2z} and the fact that 
the only dependence on $\ov Y$ is through $\Q_Y$. Now, 
from \eq{2qy}, we realize that $\,\d_{\ov Y} \,\Q_Y = 
{T\02V} \sb_P \sb_{-P}$. The two rod operators
(together with $T/V$) generate the full propagator,
hence things end up with $\, G-G=0\,$, indeed. The toy 
functions $Y$ really only reorganize a series without 
changing its meaning. 

To study the truncated free energy $F^{(\l )}\,$
it is convenient to define one more operator. Let
$\cl P_\l$ be a projector which, when applied to a
linear combination of red $g^2$ powers, suppresses all
terms $\propto g^{2m}$ with $m>\l$. Using $\cl P_\l$
we may write $F^{(\l )} = \cl E \cl P_\l F$.
The extremum of $F^{(\l )}$ has to be determined
from the equations $\d_{Y_n(P)} F^{(\l )} = 0$ with
$n=1,2,\ldots ,\l\,$. By applying these functional
derivatives to $\cl E \cl P_\l F$ one could
distinguish the following four steps. First, the 
derivatives may be interchanged with
$\cl E\,$, $\d_{Y_n(P)} \cl E
= \cl E \(\d_{Y_n(P)} + \d_{\ov Y_n (P)}\)\,$,
as in \eq{3ex}. Second, while $\d_{Y_n}$ simply
commutes with $\cl P_\l\,$, one realizes for the other
derivative, that
\be{3commu}
   \d_{\ov Y_n (P)} \;\, \cl P_\l \; = \; \cl P_{\l -n}
   \;\,\d_{\ov Y_n (P)} \quad .
\ee 
The so far reached intermediate result is
$\,\d_{Y_n(P)} F^{(\l )} = \cl E \cl P_\l \d_{Y_n(P)} F
+ \cl E \cl P_{\l -n} \d_{\ov Y_n (P)} F\,$. Third, with
view to \eq{2ser}, we note that $\d_{Y_n} F = \d_Y F$
and $ \d_{\ov Y_n} F = \d_{\ov Y} F$. But now, for the
fourth step, we may simply refer to \eq{1kap} and \eq{3ex},
to replace $\d_{Y(P)} F = T G(P)/2$ and
$\d_{\ov Y (P)} F = -TG(P)/2$, respectively. Obviously,
the two projectors form a difference$\,$:
\be{3end}
 \d_{Y_n (P)} F^{(\l )} \; = \; {T\02}\, 
 \cl E \,\( \cl P_\l  - \cl P_{\l -n} \) \; G(P) 
 \quad {\buildrel{!}\over{=}}\;\; 0 \qquad .
\ee 
Choosing $n=1$, $n=2$, and so on, it becomes obvious 
that the above condition may be equivalently
stated as $\cl E G_\l = \ldots = \cl E G_1 = 0$.
With a view to \eq{2g}, this condition reads
\begin{eqnarray}\label{3gs}
  0 &=& G_0^2 \( Y_1 - \cl E \P_1 \) \;\; ,\nonumber \\
  0 &=& G_0^2 \( Y_2 - \cl E \P_2 \)    
      + G_0^3 \(Y_1 -\cl E \P_1 \)^2 \;\; ,\nonumber \\
  0 &=& G_0^2 \( Y_3 - \cl E \P_3 \)  
    + 2\, G_0^3 \(Y_1 -\cl E \P_1 \) \(Y_2 -\cl E \P_2 \)
    + G_0^4 \(Y_1 -\cl E \P_1 \)^3   \;\; ,
\end{eqnarray} 
and so forth. Herewith we arrive at the main result of
this section. In the space of the functions $Y_1(P)$,
$\ldots$, $Y_\l(P)$, the truncated free energy $F^{(\l )}$
becomes extremal at the ''position''
\be{3min}
  \quad  Y_n = \cl E \, \P_n \qquad ,
  \qquad n = 1,\, 2,\,\ldots\, ,\, \l \quad .
\ee 
The extremum condition has obtained a formally simple
form. It is a multiple condition in the sense that one
of the $\l$ equations can be used to simplify another
(see \eq{4coco} below). Here we only remember that
the self--energy functions $\P_n$ are, of course, to be
determined diagrammatically as shown in figure 2. They
contain $Y$ through $G_0$ and $\ov Y_m$ ($m < n$)
through mass insertions. Remember that only the trivial
crosses on outer lines are absent by definition,
cf.~\eq{2g}. This explains the need for the equalizer
$\cl E$ in \eq{3min}.

%
%
\begin{figure}[bht] \unitlength1cm 
    \begin{picture}(14,1.6)
  \put(1,.1){$\P_2\; =\; - 96$}
\put(3.2,.2){\line(1,0){.8}} \put(3.6,.2){\circle{.4}}
  \put(4.4,.1){$- 144$}
\put(5.6,.1){\line(1,0){.8}}
\put(6,.3){\circle{.4}} \put(6,.7){\circle{.4}}
  \put(6.8,.1){$- 24$}
\put(7.8,.1){\line(1,0){.8}} \put(8.2,.3){\circle{.4}}
\put(8.12,.41){\x}    \put(8.04,.7){{\tiny (1)}} 
\put(9.4,.2){\vector(1,0){1}}  \put(10.9,.1){$- 96$}
\put(11.8,.2){\line(1,0){.8}}  \put(12.2,.2){\circle{.4}}
\put(12.9,.1){$= \;\P_2^{\,
  \raise 2pt\hbox{\footnotesize 2PI}}$}
\end{picture} \caption[f2]{\label{f2}{\footnotesize
   The self--energy diagrams of second order ($g^4$),
   and the two--particle irreducible part left over after
   posing the consistency condition, as explained in 
   subsection 4.2. }} \vspace{.3cm} \end{figure}
%


\let\dq=\thq \renewcommand{\theequation}{4.\dq}    
\setcounter{equation}{0}          
   
\parag{4. \ Consequences of the consistency condition}

While in the preceding section we varied $F^{(\l )}$
with respect to $Y$, we are now interested in the
value of $F^{(\l )}$ at extremum, given by \eq{3min}.
As anticipated in the section head, we shall learn
shortly, that \eq{3min} is also the right (and only right) 
consistency condition, i.e. the one that removes 
two--particle reducible diagrams, thereby much reducing
the number of terms in the perturbation series.

Irrespective of presence or absence of external legs, a
diagram is called two--particle reducible (2PR) if it
decomposes into two pieces by cutting two different
lines. Otherwise it is a two--particle irreducible (2PI)
or skeleton diagram. All diagrams in figure 1 are 2PR,
except the first and the last one, which are 2PI. The
decomposition is complete, of course$\,$:
\be{4comp}
  - \b F_n = \lk \ln \( Z_{\rm int} \) \rki_n \; = \;
  f_n \; = \; f_n^{\;\rm 2PI} \; + \; f_n^{\;\rm 2PR}
\ee  

A contribution to $f_n$ still contains $Y_m$ and $\ov Y_m$
as distinguished objects. The quantity of interest, 
instead, is $\cl E f_n$ to be taken at its extremum. The
operation of imposing the condition \eq{3min} may be 
performed in two steps. We may set $\ov Y_m = \P_m$ first, 
and specify the extremum--''position'' for $Y_m$ afterwards.
To be specific, we need a last special operator $\cl C$ 
(''the cross converter''). $\cl C$ replaces $\ov Y_n$ by 
$\P_n\,$, then $\ov Y_{n-1}$ by $\P_{n-1}\,$, $\ldots\,$,
and finally $\ov Y_1$ by $\P_1 = -12 \self\,$. Note that, 
in this ordering, also $\ov Y$ insertions (i.e. crosses)
are replaced, which occur in the $\P$ functions. 
Diagrammatically, $\cl C$ converts {\sl all} crosses 
into normal (cross--free) self--energy insertions. 
To summarize$\,$:
\be{4c}
    \lk \;\cl E \; f_n \;\rki_{\;\rm at\; extremum}
    \; = \;\lk \;\cl C \; 
    f_n \;\rki_{\;\rm at\; extremum} \;\; .
\ee 
There is no need for an equalizer $\cl E$ to the right 
of \eq{4c}, because $\cl C$ had no $\ov Y_m$ left over. 
Note that the second subscript ''at extremum'' only
applies to functions $Y$ contained in $G_0$--lines. In
the following we shall concentrate on $\,\cl C f_n\,$,
while leaving the $Y$--specification aside as some
trivial last step. To study $\cl C f_n$, all non--trivial
information can be extracted from the functional relation
\eq{1kap}. As $G_0$ carries no red $g$, this relation
holds true in each order separately,
\be{4sep} 
   2 \, G_0^{\, 2} (P) \;\d_{G_0 (P)} \; f_n 
   \; = \; G_n (P) \;\; .
\ee 
The further analysis is now decomposed into three parts
with the respective subsection head announcing the result.

\vspace{.5cm}
\under{ 4.1 \ } \ \ $\cl C \; f_n^{\;\rm 2PR} \; =\; 0 $

\vspace{.5cm}
To the left of \eq{4sep}, $f_n$ is a sum of diagrams,
and the number of $G_0$ lines in a definite diagram
may be smaller than $2n$ due to higher--order crosses
contained. The variational derivative $\d_{G_0}$,
in turn, produces one term for each $G_0$ line. 
A fixed $G_0(P)$ may occur several times at the 
same momentum argument, $q$ times, say. If $q\ge 2$,
it is called part of a ''dressed line'' (or of a
''$q$--cycle'' \cite{LW}), but if $q=1$ we call it a
''bare line''. In general, $G_0^2$ times the
differentiation with respect to a definite $G_0$ (being 
part of a $q$--cycle) will produce something with the
prefactor $G_0^{\, q+1} (P)\,$. Clearly, we may
collect all terms with the same power $q+1$ to the
left of \eq{4sep}. On the right hand side of \eq{4sep},
this grouping is already made explicit by \eq{2g}.
Each group may be identified on both sides. We learn,
that \eq{4sep} may be given a second index $q$, i.e.
it is valid separately for a given power of outer
$G_0$'s too. Here we only need the separation into
$q=1$ and $q \ge 2$. This amounts to splitting $\d_{G_0}$
into $\d_{G_0} = \d^{\rm bare} + \d^{\rm dressed}$, where 
the first derivative becomes aware of bare lines only, 
and the second notices only $G_0$'s being part of dressed 
lines. Thus$\,$:
\be{4bare}
  2\; \d^{\rm bare} \; f_n \; = \; G_0^{-2}\, G_{n,\; q=1}
     \; = \; \ov Y_n - \P_n \;\; ,
\ee 
\be{4dress}
  2\; \d^{\rm dressed} \; f_n^{\;\rm 2PR}
    \; = \; G_0^{-2}\, G_{n,\; q\ge 2} \; = \; G_0
    \sum_{m=1}^{n-1} \lk \; \rki_m \lk \; \rki_{n-m}
    \; + \; \ldots \; + \; G_0^{n-1} \lk \; \rki_1^n
\ee 
with $\lk\,\;\rki_n$ shorthand for $\, [\,\ov Y_n
- \P_n\, ]\,$. \eq{4dress} needs $n \ge 2$ (and reduction
to the last term if $n=2$). Note that there are no dressed
lines in $f_n^{\;\rm 2PI}$ by definition, but there are
bare lines in both parts, $f_n^{\;\rm 2PR}$ and
$f_n^{\;\rm 2PI}\,$. The reader (if not the journal) might
color all dressed lines in figure 1.

The functional derivative in \eq{4dress} does not change 
the ratio in which $\ov Y_m$ and $\P_m$ occur in a dressed 
line. But to the right of \eq{4dress} this ratio is the 
$\,[\,\ov Y_m - \P_m\,]\,$ combination. Hence, this
combination is just a property of dressed lines. This fact 
is well illustrated by the figure 1$\,$: the 2PR
contributions shown may be combined such that the only
remaining insertions are $\,[\,\ov Y_m - \P_m\,]\;$.
To be specific, there remain three terms$\,$:
$f_3^{\;\rm 2PR} = -3\bullet\!\!\!\bigcirc
  \hspace{-.165cm}\bigcirc\!\!\!\bullet + {1\06} \bigcirc
  \hspace{-.59cm}\lower2.3pt\hbox{$\bullet\;\bullet$}
  \hspace{-.37cm}\raise5.2pt\hbox{$\bullet$}\hspace{.23cm}
+ {1\02} \bullet\!\!\!\bigcirc
   \hspace{-.11cm}\rule[.03cm]{.14cm}{.14cm}\,$
with $\bullet \equiv \lk\,\;\rki_1$ and
$\,\rule[.03cm]{.14cm}{.14cm} \,\equiv \lk\,\;\rki_2\,$.
To check \eq{4dress} in this case, note that there is a
dressed line also in $\lk\,\;\rki_2\,$, see figure 2.

Application of $\cl C$ makes $\,[\,\ov Y_m - \P_m\,]\,$
to vanish. Any diagram in $f_n^{\;\rm 2PR}$ contains at
least one dressed line. Thus, $\cl C f_n^{\;\rm 2PR} = 0$,
and we have reached our main conclusion first, namely that
all 2PR diagrams disappear under the $\cl C$ operation$\,$,
i.e. already under the first step in posing the consistency
condition. The conclusion may be reversed. The operation
$\cl C$ is in fact the only one making $f_n^{\;\rm 2PR}$
to vanish (but let us avoid stating all arguments in
reverse order).

\vspace{.5cm}
\under{ 4.2 \ } \ \ $\cl C \; \P_n \; =\; \P_n^{\,\rm 2PI}$

\vspace{.5cm}
We return to \eq{4bare}. Its right hand side vanishes
under the $\cl C$ operation. Consequently, using
\eq{4comp},
we have
\be{4null}
  \cl C \; \d^{\rm bare} \; f_n^{\;\rm 2PI} =
   - \cl C \; \d^{\rm bare} \; f_n^{\;\rm 2PR}
  \;\; = \;\; 0 \;\; ,
\ee 
where the vanishing of the right hand side follows 
immediately from the arguments in the preceding
subsection$\,$: all dressed lines survive under
$\d^{\rm bare}$, and there is at least one in each
$f_n^{\;\rm 2PR}$ diagram. Note that $\cl C$ and
$\d^{\rm bare}$ commute, if applied to $f_n^{\;\rm 2PR}$.
But they do not, when applied to $f_n^{\;\rm 2PI}$.

To learn from \eq{4null}, we first observe that, among
the $f_n^{\;\rm 2PI}$ diagrams, there is always the blank
circle with a cross of order $n$ (the last one in
figure 1). From the rules \eq{2rules} we have
\be{4fx}
   f_n^{\,\hbox{\x}} \;\; \equiv \;\;
   \hbox{$\bigcirc$\raise 6pt\hbox{\hspace{-.3cm}\x}
   \raise 14pt\hbox{\hspace{-.37cm}\tiny ($n
   $)\hspace{.3cm}}} = \; \b\, V \,\sum_P {1\02}
   \ov Y_n(P) \, G_0(P)  \quad , \quad
   2\,\cl C \; \d^{\rm bare} \; f_n^{\,\hbox{\x}} 
   \; = \;\cl C \; \ov Y_n  \; = \;\cl C \, \P_n \;\; .
\ee 
Now, using \eq{4fx}, we may rewrite \eq{4null} as
\be{4pi}
  \cl C \,\P_n = -2\;\d^{\rm bare}\; 
  f_n^{\,\prime\;{\rm 2PI}}  \qquad \hbox{with} \qquad
  f_n^{\,\prime\;{\rm 2PI}} \equiv f_n^{\;\rm 2PI}
     - f_n^{\,\hbox{\x}} \;\; .
\ee 
Note that there are neither dressed lines nor crosses 
in $f_n^{\,\prime\;{\rm 2PI}}$ diagrams, by definition. The
same happens with $\d^{\rm bare} f_n^{\,\prime\;{\rm 2PI}}$.
Therefore we dropped one $\cl C$ in \eq{4pi}. 
$\d^{\rm bare} f_n^{\,\prime\;{\rm 2PI}}$ is made up of
diagrams having two ends. If cutting two inner lines of
such a diagram, it remains connected. So, it contributes
to $\P_n^{\,\rm 2PI}$. Moreover, these contributions form
$\P_n^{\,\rm 2PI}$ itself, because through the above
reformulations we only removed diagrams. This argument
makes \eq{4pi} to become
\be{4like}
  \cl C \,\P_n  \;\; = \;\; \P_n^{\,\rm 2PI} \qquad ,
  \qquad n = 1,\, 2,\,\ldots\, ,\, \l \quad .
\ee 
For the simplest example see figure 2. For $n=1\,$, 
admittedly, \eq{4like} is trivial, since $\P_1^{\,\rm 2PI}
= \cl C \P_1 = \P_1 = -12 \self\,$. The result \eq{4like} 
is very welcome for a final simplification of the 
consistency condition \eq{3min}$\,$:
\be{4coco}
  Y_1 = \P_1 \lk Y \rk  \;\;\; , \;\;\;
  Y_2 = \P_2^{\,\rm 2PI} \lk Y \rk \;\;\; , \;\;\;
  \ldots \;\;\; , \;\;\; Y_\l = \P_\l^{\,\rm 2PI}
  \lk Y \rk \;\;\; ,
\ee 
where the functional dependence on $Y=Y_1 + \ldots + Y_\l$
is hidden in the bare $G_0$ lines. Note that \eq{4coco} 
resulted from an iterative use of the multiple condition 
\eq{3min}, thereby minimizing the number of 
sum--integrals involved. Of course, \eq{4coco} may be
written as a single equation, $Y=\sum_{n=1}^\l \P_n^{\,
\rm 2PI}\,[\, Y\,]\,$. But note, that only with a finite
number $\l$ of terms, it makes sense to be a generalized
gap equation.

\vspace{.5cm}
\under{ 4.3 \ } \ \ $\cl C \; f_n^{\;\rm 2PI} \; = \;
                     (1-2n)\; f_n^{\prime \;\rm 2PI}$

\vspace{.5cm}
We now concentrate on the $n$-th order diagrams remaining 
after all the above reductions. To start with, we remember
\eq{4comp}, $\cl C f_n^{\;\rm 2PR}=0$ and the definition
in \eq{4pi} to get
\be{4fs}
  \cl C \; f_n \; = \;  \cl C \; f_n^{\;\rm 2PI}
  \; = \; \cl C \; f_n^{\,\hbox{\x}}
  \; + \; f_n^{\prime \;\rm 2PI}
\ee 
with $f_n^{\prime \;\rm 2PI}$ to be obtained from the rules
\eq{2rules} by omitting all dressed and/or crossed
diagrams. Further such contributions will arise, if
(by $\cl C$) the $n$-th order cross in $f_n^{\,\hbox{\x}}$
is replaced by $\P_n^{\;\rm 2PI}$. So, there could be a
relation between the two terms to the right in \eq{4fs}.
This happens indeed, and is easily established by combining
the equations \eq{4fx}, \eq{4pi} and \eq{4like}$\,$:
\be{4mit}
  \cl C f_n^{\,\hbox{\x}} \; = \; \b\, V \;\sum_P
     {1\02} G_0(P) \P_n^{\;\rm 2PI}(P) \; = \;
     - \b\, V \; \sum_P G_0 (P)
  \d_{G_0(P)} \; f_n^{\;\prime\;{\rm 2PI}}  \;\; .
\ee 
Remember that $\d_{G_0}$ removes a sum together with the 
factor $\b V$. But precisely these details are restored 
on the above right hand side. The operator $\b V\sum G_0 
\d_{G_0}$ just counts lines. There are $2n$ lines in each 
diagram of $f_n^{\;\prime\;{\rm 2PI}}$, and that is it$\,$:
$\cl C f_n^{\,\hbox{\x}} = - 2n f_n^{\;\prime\;
{\rm 2PI}}\,$. To summarize, the element $\cl C \, f_n$,
which is needed on the right hand side of \eq{4c}, is
given by $\,\cl C f_n = \cl C f_n^{\;\rm 2PI} = (1-2n)
f_n^{\;\prime\;{\rm 2PI}}\,$, as announced.

The free energy, consistently resummed up to order
$\l\,$, is thus obtained as
\be{4final}
   \lk \; - \,\b\, F^{(\l)} \; \rki_{\;{\rm at\; extremum}}
   \;\;\; = \;\;\;
   \Big[ \;\; f_0 \; + \;\sum_{n=1}^\l (1-2n)\,
         f_n^{\;\prime\;{\rm 2PI}} \;\;\;
   \Big]_{\; Y=\sum_{n=1}^\l \P_n^{\,{\rm 2PI}} \lk Y \rk}
\ee 
or, equivalently, from the following set of diagrammatic
rules$\,$:
\be{4rul}   \hspace{.2cm}
  \parbox{14cm}{{\footnotesize
1. Drop all remarks on crosses in the rules 
   \eq{2rules}$\,$, $\,$i.e. \hfill \\[-1pt]
   \hbox{\hspace{.35cm} return to} the rules for
   $\phi^4$ theory without toy mass. \hfill \\[-1pt]
2. Draw 2PI diagrams only. Come to a decision for
   the truncation$\,$: $n \le \l\,$.  \hfill \\[-1pt]
3. Multiply each combinatoric factor of an $n$-th order 
   diagrams with $(1-2n)\,$. \hfill \\[-1pt]
4. To specify the function $Y=Y_1 + Y_2 + \ldots + Y_\l $ 
   in $G_0\,$, solve the gap equations \eq{4coco}. \hfill
}} \ee 

The result \eq{4rul} is agreeably simple. As figure 3
shows, the number of diagrams has reduced so much, that
the remainder can be presented up to $g^{10}$ with ease.
The combinatoric factor for pearl rings, given in the
caption, persists to higher orders and can be proven by 
induction. Once \eq{4final}, \eq{4rul} are reached, the 
details on cross insertions may be viewed as some less 
important aspect of the derivation. To avoid possible
misunderstandings (taken up in the next two paragraphs),
let us emphasize the generality of the result and recall
its meaning. The ingoing question (for the correct
decomposition of the Lagrangian whose truncated
perturbation series gives a well--organized asymptotic
expansion) is answered by the gap equation, i.e. by the
subscript to the right of \eq{4final}. But \eq{4final}
itself merely states the result of working with the
solution $Y(Q)$.

%
%
\begin{figure}[bht] \unitlength1cm 
    \begin{picture}(16,3.7)
  \put(.56,2.6){$ -\,\b\, F^{(5)} \;\; = \;\; {1\02}
       \hspace{.62cm} - \, 1\cdot 3$}
  \put(3.6,2.76){\circle{.4}}           
\put(5.56,2.76){\circle{.4}\circle{.4}} 
  \put(6.26,2.6){$- \; 3\cdot 12$} 
\put(8.0,2.76){\circle{.4}}  \put(8.2,2.76){\circle{.4}}
  \put(8.55,2.6){$- \; 5\cdot 2\cdot 12^2$} 
\put(10.95,2.7){\circle{.4}\circle{.4}}
  \put(11.15,3.04){\circle{.4}}
  \put(11.7,2.6){$- \; 7\cdot {3\02}\cdot 12^3$} 
\put(14.15,2.6){\circle{.4}\circle{.4}} 
\put(14.15,3){\circle{.4}\circle{.4}}
  \put(4,1.4){$- \; 7\cdot 6\cdot 12^3$} 
\put(6.5,1.5){\circle{.6}}  \put(6.93,1.5){\circle{.32}} 
\put(6.81,1.5){\circle{.6}}
  \put(7.3,1.4){$- \; 9\cdot {6\05} \cdot 12^4$} 
\put(9.8,1.3){\circle{.3}\circle{.3}}
\put(10.2,1.6){\circle{.3}}  
\put(9.7,1.6){\circle{.3}}  \put(9.95,1.8){\circle{.3}}
  \put(10.6,1.4){$- \; 9\cdot 12 \cdot 12^4$} 
\put(13.25,1.4){\circle{.4}} \put(13.45,1.4){\circle{.4}}
\put(13.22,1.72){\circle{.25}\circle{.25}}
  \put(4,.2){$- \; 9\cdot 24 \cdot 12^4$} 
\put(6.812,.3){\circle{.6}} \put(7.388,.3){\circle{.6}}
\put(7.1,.3){\circle{.7}}
  \put(8,.2){$- \; 9\cdot {16\05} \cdot 12^4$} 
\put(11.3,.34){\oval(1.5,.9)}
\put(11.3,-.06){\oval(1.04,1.02)[t]}
\put(11.3,.26){\oval(.86,1.04)[t]}
\put(10.78,-.06){\line(3,1){.94}}
\put(11.8,-.06){\line(-3,1){.4}} 
\put(10.86,.256){\line(3,-1){.36}} 
\end{picture} \caption[f3]{\label{f3}{\footnotesize
   The contributions to the resummed free energy up
   to 5--th order ($g^{10}$). The combinatoric
   factors are written as products with $(1-2n)$ the 
   first factor of each. For rings of pearls (4., 5.
   and 7. diagram) the combinatoric factors follow the
   rule $(1-2n)\, 12^n/(2n)\,$. The last diagram is
   non--planar. The first term is $-\b F_0$, and its
   factor denotes the number of blackbody radiations
   at $Y\to 0$, cf. \eq{4lim}.
   }} \vspace{.3cm} \end{figure}
%

By \eq{4final} one could be strongly remembered to
structures observed by Cornwall, Jackiw and Tomboulis
\cite{cjt} in their study of the effective potential. Gap
equations mark stationary points and reduce the effective
action diagrams to skeletons with lines being improved
propagators \cite{cjt,camel,nbag}. It may happen that
\eq{4final} can be derived along these lines as well
(though the diagrams shown there are of rather low
order). We have not followed up this possibility. But
we are able to perform a near--by other test next.

{}From even earlier days \cite{lutt,bloch,freed,LW} one
knows of the possibility of {\sl exactly} summing
self--energies into the lines of skeleton diagrams. But
remember that, in the absence of truncation, the terms
consistency and gap equation make no sense, and that the
exact free energy was realized in \eq{3ex} to be a merely
uninteresting limiting case. Nevertheless, we are now
invited, to test \eq{4final} by taking the limit $\l \to
\infty\,$. The condition \eq{4coco} becomes
$Y = \sum_{n=1}^\infty \P_n^{\,\rm 2PI}\,[\, Y\,]\,$ and
has the solution $Y=\P$, with $\P\equiv\sum_{n=1}^\infty
\P_n\,[\,\ov Y =Y=0\,]\,$ the exact self--energy,
cf.~\eq{2g}. Next, concerning the $-2n$ part
of the third rule, we step back to $f_n^{\,\hbox{\x}}\,$:
$-\b F^{(\infty )} = f_0 + \sum_{n=1}^\infty
f_n^{\prime\;{\rm 2PI}} + \sum_{n=1}^\infty \cl C
f_n^{\,\hbox{\x}}\,$. Using \eq{4mit}, we may perform the
second sum. Note that $Y=\P$ makes $G_0(P) = 1/(\P(P)
- P^2)$ to be the exact propagator of the original
($Y$--free) theory. In presenting $f_0\,$ we care about
the correct behavior $f_0\to VT^3\pi^2/90$ at vanishing
coupling. Then, from Appendix A of \cite{vari}, we are
led to introduce the function
$r(P) \equiv T^2 e^{\b\vert {\bf p}\vert} \,\d_{n,\, 0}
- P_0^2 \;$ (remember $P_0^2 = - T^2 (2\pi n)^2\,$). 
Herewith, the present exercise ends up with 
\be{4lim}
  - {1\0V}\; F^{(\infty )} \; = \; {1\02}\;
  \sum_P \,\ln \( { r(P) \0 \P(P) - P^2 } \)
  \; + \; {1\02} \;\sum_P \; { \P(P) \0 \P (P) - P^2 }
  \; + \; {1\0\b V} \;
   \sum_{n=1}^\infty f_n^{\prime\;{\rm 2PI}} \;\; ,
\ee 
which (for $r=1\,$) is equation (22.16)
in \cite{bloch}, or (4.2.10) in \cite{LW}. We have thus
rederived the old exact statement, indeed.


\let\dq=\thq \renewcommand{\theequation}{5.\dq}    
\setcounter{equation}{0}          
   
\parag{5. \ The first few terms}

and the one--loop gap equation. Although looking quite
different, our resummed free energy must precisely
reproduce the asymptotic expansion for the $\phi^4$ pressure,
as far as known. We shall be content here to demonstrate
this reproduction. For this task, we may concentrate on
$F^{(2)}\,$, given by the first three diagrams of figure 3,
as well as on the gap equation \eq{4coco} for $\l =2\,$.

Under renormalization several quantities change their
meaning. The Lagrangian \eq{1la} becomes the renormalized
one and $g$ turns to be the running coupling. Among the
counter terms, only $Z_2 = 1 + 3g^2/(32\pi^2 \e) + O(g^4)$
is of relevance \cite{az,zk}, where $\e$ refers to
dimensional regularization$\,$: $\sum_P \to T \sum_n
\mu^{2\e} (2\pi)^{2\e-3}\int d^{3-2\e} p$. The
$\phi^4$-interaction in \eq{1la} modifies by $g^2 \to
\mu^{2\e} Z_2 g^2$. The corresponding $g^4$ term must be
included in the second diagram of figure 3 and in the
first of the conditions \eq{4coco}. The second condition
\eq{4coco}, may be simplified immediately$\,$: $Y_2(Q) =
- {1\06} g^4 \sum_{P,\, K} G_0 (P) G_0 (K) G_0 (P+K+Q)\,$
with the crude approximation $Y \approx g^2 T^2/24$
sufficient in $G_0$.

The first condition \eq{4coco} needs more care. Since
$Y_1$ does not depend on the outer momentum $Q$ (a fact
special to $\phi^4$ theory), we may write
$Y_1 \equiv m^2 + \d$, i.e.
\be{5c}
   m^2 + \d = {1\02} g^2 \sum_P {1\0 m^2 - P^2
   + \d + Y_2} + {1\02} g^2 (Z_2-1) \sum_P {1\0 m^2
   - P^2 + \d + Y_2} \;\; ,
\ee 
and determine $\d$ such that the leading part of
\eq{5c}, the one with $G_{00} \equiv 1/(m^2-P^2)$ under 
the first sum, turns into the one--loop gap equation
\be{5true}
  m^2 \; = \; {1\02} \, g^2 \,\sum_P\,
  {1\0 m^2 - P^2} \quad .
\ee 
Expanding \eq{5c} one obtains
$\;\d = {1\02} g^2 (Z_2-1) \sum_P G_{00} - {1\02} g^2
  \sum_P G_{00}^2 \( \d + Y_2 \) + O(g^7)\,$.

For the pressure $-F^{(2)}/V \equiv p$ we have three
contributions, $p = p_0 + p_1 + p_2$, corresponding to
the first three diagrams of figure 3$\,$: $p_0 = {1\02}
\sum_P \ln ( r G_0 )$, cf. \eq{4lim}, $p_1 = {1\08}
Z_2 g^2 (\sum_P G_0 )^2$ and $p_2= - {1\016} g^4
I_{\rm ball}$ (overall factors $\mu^{-2\e}$ are omitted
for brevity). As with \eq{5c}, we might collect those
leading terms of $p_0+p_1$, called $p_m$, which contain
$G_{00}$ and require a nontrivial solution of \eq{5true}
for $m$. Hence, $p = p_m + p_{\rm rest}$. After a bit of
calculation we obtain
\be{5mr}
  p_m = {1\02} \sum_P \, \ln ( r\, G_{00} )
    + {g^2\08}\, (\sum_P G_{00})^2 \;\; , \;\;
  p_{\rm rest} = - {g^2 (Z_2-1) \0 8}\, (\sum_P G_{00})^2
    + {g^4 \0 48} I_{\rm ball} 
\ee 
with $\, I_{\rm ball} = \sum_{Q,\, P,\, K} G_0 (Q) G_0 (P)
G_0 (K) G_0 (Q+P+K)\,$ \cite{az}. Two terms proportional
to $(\d + Y_2)$ have canceled in $p_{\rm rest}$, because
$p$ is at its minimum. Again, being content with $\le g^5$,
the functions $G_0$ in $p_{\rm rest}$ may be supplied with
the lowest order value of $m^2$. Note that, with $p_{\rm
rest}$, two of the five contributions of ref. \cite{az} are
obtained, namely those of fourth order in red $g$'s. Hence,
the remaining three diagrams, which are $\bigcirc$,
$\bigcirc\hspace{-.08cm}\bigcirc$ and
$\bigcirc\hspace{-.13cm}$\raise 1pt\hbox{\x}$\,$, might
be anyhow hidden in $p_m$. But note that, in these three
diagrams, cross and lines carry the Arnold--Zhai value
$m_A^2$, which is given by \eq{5true} at zero mass to the
right. We are thus led to expand $p_m$ around $m_A^2\,$:
\be{5ex}
  p_m = \lk p_m \rki_A - {1\02} (m^2-m_A^2)\,\Big[\,
  1 - {1\02} g^2 \6_{m_A^2} \sum_P G_{00\; A}\,\Big]
  \,\sum_P G_{00\; A} + O(g^5) \;\; ,
\ee 
where $G_{00\; A} \equiv \sum_P 1/(m_A^2-P^2)$ and,
at this point, we relax retaining $g^5$ terms.
They need no additional diagram \cite{pasi}. Now, 
the gap equation \eq{5true}, if expanded around 
$m_A^2\,$,
\be{5a}
   \( m^2 - m_A^2 \) \; \Big[\, 1 - {1\02} g^2
   \6_{m_A^2} \sum_P G_{00\; A}\,\Big] \; = \;
   {1\02}\, g^2 \sum_P G_{00\; A} - m_A^2 \;\; ,
\ee 
is seen to be of direct use in \eq{5ex}. Moreover, it
gives $p_m$ the desired form to exhibit the three diagram
contributions in search. In total, the terms obtained
combine to the well known expression (in our language)
\be{5tot}
   p = {1\02} \sum_P \ln \( r \, G_{00\; A} \) 
   - {1\08}\, Z_2\, g^2 \Big( \sum_P G_{00\; A} \Big)^2
   + {1\02}\, m_A^2 \sum_P G_{00\; A}
   + {1\0 48}\, g^4\, I_{\rm ball} \;
\ee 
for the pressure up to three--loop order. At these low 
orders, and already on an algebraic level, the resummed 
theory has reduced to the traditional setup, indeed.

It might have been remarked, that our low order limit
has led to the Arnold and Zhai version \cite{az} of 
the asymptotic expansion, but not to that of Parwani 
and Singh \cite{pasi}, whose mass term includes 
$\d_{P_0, 0}\,$. But the latter version only amounts to
a regrouping of terms having the same order of magnitude
\cite{AE}. Note that the absence of ''forbidden'' $g^3$
terms in equation (12) of \cite{pasi} is still due to a
suitable choice of the toy mass prefactor.

Through all of the preceding sections (starting with the
example \eq{1az}), it was taken for granted, that (a) 2PR
diagrams reduce the $g$ order, hence being forbidden in a
systematic asymptotic expansion, and that (b) there is no
other mechanism producing $g^{-1}$ factors. Concerning
statement (a), consider, without loss of generality, a
diagram with only one dressed line, a $q$-cycle with
$q=k+1$ and $k\ge 1$. With $m^2 \sim g^2$ a constant part
of $Y$ and $\6_m = 2 m \,\6_{m^2}$ we may write
\be{5con}
 \sum_P G_0^{k+1} (P) h(P) = \sum_P
 {h(P) \0 k!} (-\6_{m^2} )^k G_0 (P) =
  {1\0 k!} \( - {1\0 2m}\,\ov{\6}_m \)^k \sum_P h(P)
  G_0 (P) \;\; ,
\ee 
where $h(P)$ is a $(k+1)$-fold product of 2PI self--energy
functions (or cross insertions) and $\ov{\6}$ is not
allowed to act on the $m$'s in $h(P)$. Even under this
restriction, all experience with evaluated skeleton 
diagrams $\sum h(P) G_0(P)$ \cite{az,pasi} shows, that 
they have a term $\propto\! m$ in its asymptotic expansion.
But this is sufficient to reduce its order through
$\,(- {1\0m} \6_m )^k m = - (2k-3)!! \; m^{1-2k}\,
\sim g^{-2k+1}\,$. The statement (b) is somewhat delicate
as we have no proof for. But it is hard to realize, that
the $g$ order could be reduced anyhow else than by
\eq{5con}. Statement (b) is the strong conjecture, this
paper rests on \ --- \ and ends up with.


\parag {6. \ Conclusions }

To summarize, the small--coupling asymptotic expansion
for the $\phi^4$ thermodynamics is supplied with
a general consistency condition. The latter is derived
by requiring the free energy to be extremal, but then
shown to guarantee the systematics of the asymptotic
expansion. By the corresponding resummation, the
pressure is given by simple diagrammatic rules. But
the self--energy in their ''bare'' line propagators
need to be self--consistently determined by solving
a generalized gap equation. Former treatments are
demonstrated to be low--order special cases of this
scheme.

Most probably, the observed structure has its counterpart
for gauge theories as well, at least with regard to the
functional methods used in this paper. The hot Yang--Mills
system (pure gluon plasma) is under present study.

%
\vspace{.5cm} \par\noindent
\parbox[t]{13cm}{\hspace{.84cm}
{\bf Acknowledgments}\\[4pt] {\sl
We are grateful to Marc Achhammer, Fritjof Flechsig
and York Schr\"oder \\ for valuable discussions.}}


\let\dq=\thq \renewcommand{\theequation}{A.\dq}
\setcounter{equation}{0}          
   
\parag{Appendix}

Here we comment on the generation of diagrams while
retaining the information on combinatoric factors.
Diagrams derive from \eq{2z} by first expanding the
exponential. Imagine $n$ operators $\Q$ applied to
$W_0\,$. Reduce $\Q_Y$, \eq{2qy},  and $\Q_g$, \eq{2qg},
to the rod operators, because the other details ($\sum$,
weight, Kronecker) are preserved by the rules \eq{2rules}.
Now move $W_0\,$ to the left by commuting it with each
$\sb\;$, $\sb\; W_0 = W_0\, ( \sb + j )\,$, where $j$ is
shorthand for $G_0 \schl \jmath$ (but $G_0$ is preserved
by the rules). At the very left $W_0$ may be omitted due
to the $j=0$ prescription in \eq{2z}. Let $\6$ be a
rod--derivative, which is not allowed to act on a $j$ on
the same vertex. These inner differentiations may be made
explicit, instead. After all this, $\Q$ has converted to
\be{a1}
 {\bf D} \;\equiv\;
   6 \propto  + 12 \propto^{\6} +\, 6 \propto^{\6}_{\6}
   + \times +\, 4 \times^{\6} +\, 6 \times^{\6}_{\6}
 +\, 4 \;\times^{\!\6}_{\hspace{-.45cm}\6\hspace{.24cm}\6}
 + \;\;\lower.11cm\vbox{\hbox{$_{\6} \hskip .246cm _{\6}$}
   \vskip -.34cm \hbox{\hskip .15cm ${\dis \times}$}
   \vskip -.5cm \hbox{$_{\6} \hskip .246cm _{\6}$}}\;\;
   + \,\vert\!\!\hbox{\x} 
   + 2 \;\vert\!\!\hbox{\x}
    \hspace{-.15cm}\rule[-.3cm]{0cm}{.77cm}^\6 
   + \,\vert\!\!\hbox{\x}
    \hspace{-.15cm}\rule[-.3cm]{0cm}{.77cm}_\6^\6 \;\; .
\ee 
Unspecified line ends carry $j$. An unspecified cross
represents the sum $\sum_{k=1}^\infty$ over $k$-th order 
crosses. Two terms, $3\infty$ and 
$\bigcirc\hspace{-.12cm}$\raise1pt\hbox{\x}$\,$, are 
omitted in \eq{a1} since they certainly lead to
disconnected diagrams. Now, form $(1/n!)\;{\bf D} \ldots
{\bf D}\; 1\;$, drop further disconnected pieces and set
$j=0$, finally, i.e. omit diagrams with empty ends. Having
obtained a definite diagram, its combinatoric factor may
be checked, of course, against $-(4!)^v/(2mS)$ with $v =$
number of vertices, $m=$ multiplicity of a line cut up,
$S^{-1}=$ symmetry factor of the self--energy diagram
arisen.

Up to $n=3$, the above {\bf D} operations may be well 
performed by hand (figure 1). But let Miss MAPLE continue
to higher orders (figure 3). The program must be given
some memory for {\sl which} two vertices have been joined
by a line. The corresponding crucial program--line reads
\hbox{$\;$w:=proc(b,a,x); b.a*diff(x,a); end;$\;$}.


    \vspace{.2cm}   \renewcommand{\section}{\paragraph}

\vspace{.7cm}  \centerline{\ \ \ $\,\bigcirc
    $\hspace{-.92cm}\rule[2.9pt]{1.4cm}{.2pt}}
\end{document}